\documentclass[aps,pra,twocolumn,floatfix,showpacs,noshowkeys]{revtex4}%
\usepackage{amsmath,amssymb,bm,color,graphicx,epsfig,graphics}

\newcommand{\ket}[1]{|{#1} \rangle}
\newcommand{\bra}[1]{\langle {#1}|}

\begin{document}
\preprint{MPI-PKS}
\title{Quantum Computational Method of Finding the Ground State Energy 
       and Expectation Values}
\author{Sangchul Oh}~\email{scoh@pks.mpg.de}
\affiliation{Max Planck Institute for the Physics of Complex Systems,
             N\"othnitzer Str. 38, D-01187, Dresden,
             Germany}
\date{\today}
\begin{abstract}
We propose a new quantum computational way of obtaining a ground-state energy 
and expectation values of observables of interacting 
Hamiltonians. It is based on the combination of the adiabatic quantum 
evolution to project a ground state of a non-interacting Hamiltonian onto 
a ground state of an interacting Hamiltonian and the phase estimation 
algorithm to retrieve the ground-state energy. The expectation value 
of an observable for the ground state is obtained with the help of 
Hellmann-Feynman theorem. As an illustration of our method, we consider 
a displaced harmonic oscillator, a quartic anharmonic oscillator, 
and a potential scattering model. The results obtained by this method 
are in good agreement with the known results.
\end{abstract}
\maketitle

\section{Introduction}

Quantum simulation might be a real application of medium-scale quantum 
computers with $50-100$ qubits~\cite{Schack06}. As Feynman suggested, 
a quantum computer can simulate quantum systems better than a classical 
computer because it is also a quantum system~\cite{Feynman82}. 
Lloyd demonstrated that almost all quantum systems can be simulated on 
quantum computers~\cite{Lloyd96}. Abrams and Lloyd presented a quantum 
algorithm to find eigenvalues and eigenvectors of a unitary operator based on 
the quantum phase estimation algorithm~\cite{Abrams99}. 
Although it is an efficient quantum algorithm, there is room for improvement. 
First, one has to prepare an input state close to unknown 
eigenstates. Second, it has been little explored 
how to obtain physical properties except the energy spectrum.

In this paper, we propose a new refined quantum computational method to 
calculate the ground state energy and expectation values of observables 
for interacting quantum systems. The main idea is as follows. Adiabatic 
turning on an interaction makes the ground state of a non-interacting 
system evolve to the ground state of an interacting system. During 
the adiabatic evolution, the phase estimation algorithm extracts 
the phase of an evolving quantum system continuously without the collapse 
of a quantum state. So the ground energy of an interacting system is 
obtained as a function of coupling strength. With the help of 
the Hellmann-Feynman theorem~\cite{Hellmann}, the expectation value of
an observable for the ground state of an interacting system is obtained. 
As a test of our method, we simulate on classical computers three quantum 
systems: a displaced harmonic oscillator, a quartic anharmonic 
oscillator~\cite{Bender69}, and a potential scattering model~\cite{Kehrein}.

\section{Method}

Let us start with a brief review of Abrams and Lloyd's algorithm. 
Its goal is to find eigenvalues $E_n$ and eigenstates $\ket{E_n}$ of 
a time-independent Schr\"odinger equation 
\begin{align}
H \ket{E_n} = E_n \ket{E_n} \,.
\label{time_indep_schroedinger}
\end{align}
Their key idea to solve (\ref{time_indep_schroedinger}) is to consider its
time evolution 
\begin{align}
e^{-iHt/\hbar}\,\ket{\Psi_{I}} =  \sum_{n=0} e^{-iE_nt/\hbar}\,a_n \ket{E_n}\,,
\label{time-evolution}
\end{align}
where $\ket{\Psi_I} =\sum_{n} a_n\ket{E_n}$ is an input or trial state.
The information on eigenvalues $E_n$ in the input state is 
transfered to index qubits by applying the quantum phase estimation algorithm.
The measurement of the index qubits gives us a good 
approximation to $E_n$ with probability $|a_n|^2$, and makes $\ket{\Psi_I}$ 
collapse to $\ket{E_n}$. It is instructive to compare (\ref{time-evolution}) 
with the quantum Monte Carlo method which uses the imaginary time $\tau = it$ 
to project the input state onto the ground state~\cite{Foulkes01} 
\begin{align}
\lim_{\tau\to\infty} e^{-H\tau/\hbar}\,\ket{\Psi_I} 
\simeq e^{-E_0\tau/\hbar}\,a_0 \ket{E_0}\,.
\label{QMC}
\end{align}
First, in order to find the ground state energy, both (\ref{time-evolution}) 
and (\ref{QMC}) require a good input state close to $\ket{E_0}$. 
If the input state does not contain the information about the ground 
state, both will fail. Second, for each run, while (\ref{time-evolution}) 
outputs $E_n$ randomly, (\ref{QMC}) produces $E_0$ always. Finally, 
(\ref{time-evolution}) is a real time evolution, however, (\ref{QMC})
is the imaginary time evolution, i.e.,  a diffusion process, which 
is implemented by classical random walks. 

Our goal is to find a ground state energy with probability 1 even 
if an input state contains little information on the ground state.
Our method uses a real time projection onto the ground state by 
adiabatically turning on an interaction. Ortiz {\it et al.} suggested 
the use of the Gell-Mann-Low theorem to find the spectrum of a
Hamiltonian with quantum computers~\cite{Gellmann51,Ortiz01}. 
Farhi {\it et al.} developed the adiabatic quantum computation~\cite{Farhi01}.

We divide the Hamiltonian $H$ into two parts: non-interacting Hamiltonian 
$H_0$ and interaction $H_1$, $H = H_0 + H_1$, As usual, it is assumed that 
the eigenvalues $W_n$ and eigenstates $\ket{W_n}$ of $H_0$ are known, 
$H_0\ket{W_n}= W_n \ket{W_n}$. We recast $H$ to be time-dependent 
\begin{align}
H(t) &= H_0 + f(t)H_1  + E_c\,,
\label{time_dep_Hamil}
\end{align}
where a slowly varying function $f(t)$ satisfies 
$f(0) = 0$ and $f(T_R) = 1$ with running time $T_R$. 
The role of the constant energy $E_c$ will be
explained later. As the interaction is turned on slowly, the input state 
$\ket{W_0}$ evolves adiabatically to $\ket{E_0}$
\begin{align}
\hat{T}\,e^{-\frac{i}{\hbar}\int_0^t\, H(t')\,dt'}\,\ket{W_0} 
\simeq e^{-\frac{i}{\hbar}\int_0^t\, E_0(t') dt' }\, \ket{E_0} \,,
\label{adiabatic-evolution}
\end{align}
where $\hat{T}$ is a time-ordering operator~\cite{Geomtric}. 
Notice the similarity and difference between (\ref{time-evolution}),
(\ref{QMC}), and (\ref{adiabatic-evolution}). 
The quantum phase estimation algorithm can extract the information 
on $E_0$ from (\ref{adiabatic-evolution}). 
Since, during the adiabatic evolution, the quantum system is in an 
instantaneous ground state $\ket{E_0(t)}$ of $H(t)$, one can apply 
frequently the phase estimation algorithm without the collapse 
of the quantum state to the exited states. 

Since the phase $\phi= E_nt/\hbar$ is defined in $0\le \phi < 2\pi$, 
the phase estimation algorithm gives us only the absolute value of an 
energy $|E_n|$. Its sign can be determined by adding $E_c$. 
When $E_0$ is negative, while $W_0$ is positive, 
$E_c>E_0$ makes all the spectrum positive. 
Also $E_c$ is useful for stabilizing the algorithm. 
If $|E_0|$ is close to zero, a long time needs to make the phase 
$\phi =E_0t/\hbar$ finite.

The expectation value of an observable ${\cal O}$ can be obtained with 
the help of the Hellmann-Feynman theorem~\cite{Hellmann}. It states that 
if $H(\alpha) \ket{E_n(\alpha)} = E_n(\alpha) \ket{E_n(\alpha)}$ with 
parameter $\alpha$, then the following relation holds 
\begin{align}
\frac{dE_n(\alpha)}{d\alpha} 
= \bra{E_n(\alpha)} \frac{dH}{d\alpha} \ket{E_n(\alpha)}\,.
\label{Hellmann_Feynman}
\end{align}
By modifying the full Hamiltonian to have a linear coupling to ${\cal O}$, 
$H(t) = H_0 + f(t)(H_1 + \alpha {\cal O}) + E_c$, (\ref{Hellmann_Feynman})
becomes 
\begin{align}
\left.\frac{dE_n(\alpha)}{d\alpha}\right|_{\alpha = 0} 
= \bra{E_n} {\cal O} \ket{E_n}\,.
\label{HF_Numeric}
\end{align}
Therefore, the expectation value of an observable is obtained from
a derivative of $E_n(\alpha)$ at $\alpha = 0$. In practice, (\ref{HF_Numeric}) 
is obtained from a numerical approximation $ \bra{E_n} {\cal O} \ket{E_n}=
\left[E_n(\alpha) - E_n(-\alpha)\right]/2\alpha + {\cal O}(\alpha^2)$.
This is comparable with an expectation estimation algorithm~\cite{Knill07}.
Notice that our scheme does not require the repeated measurements and the average over 
the individual outcomes~\cite{Schack06}.

\section{Application to Quantum Systems} 
\subsection{Displaced harmonic oscillator} 

As an illustration of our method, let us consider a simple Hamiltonian,
\begin{align}
H_0 = \frac{p^2}{2m} + \frac{m\omega^2 x^2}{2}\,,
\quad H_1 = \lambda x \,.
\label{DHO}
\end{align}
For convenience, we set $\hbar = m = \omega = 1$. It is well known that
(\ref{DHO}) is exactly solvable, a usual perturbation theory for it 
works well, and its ground state is a coherent state. 

The first step to quantum simulation is to map a physical system to 
a qubit system. The position $x$ in~(\ref{DHO}) is continuous, but 
qubits are discrete.  A usual approach is to discretize $x$. 
Another way is to map the eigenstates $\ket{n}$ of $H_0$ to 
the computational basis of $N$ qubits,
$\ket{n} = \ket{j_{N-1}}\otimes\ket{j_{N-2}}\otimes\dots\ket{j_0}$ 
with $n = j_{N-1}\,2^{N-1} + j_{N-2}\,2^{N-2} + \dots + j_{0}\,2^0$
and $j_k = 0$ or $1$. Then $H_0$ is given by a diagonal matrix,
\begin{align}
H_0 \approx \sum_{n=0}^{2^N-1}\left(n+\frac{1}{2}\right)\ket{n}\bra{n} \,.
\label{DHO_0}
\end{align}
The quantum dynamics of (\ref{DHO_0}) was simulated on 
an NMR quantum computer by Somaroo {\it et al.}~\cite{Somaroo99}. 
On the other hand, $H_1$ is written as a tridiagonal matrix,
\begin{align}
H_1 \approx \frac{1}{\sqrt{2}} \sum_{n=0}^{2^N-2} 
    \left(\, \sqrt{n}\,\ket{n}\bra{n+1} 
	  + \sqrt{n}\,\ket{n+1}\bra{n}\, \right) \,.
\label{DHO_H1}
\end{align}
A quantum state $\ket{\psi(t)}$ at time $t$ can be expressed in terms of
$\ket{n}$, $\ket{\psi(t)} = \sum_{n=0}^{2^N-1} a_n(t) \ket{n}$.

\begin{figure}[htbp]
\includegraphics[scale=1.0]{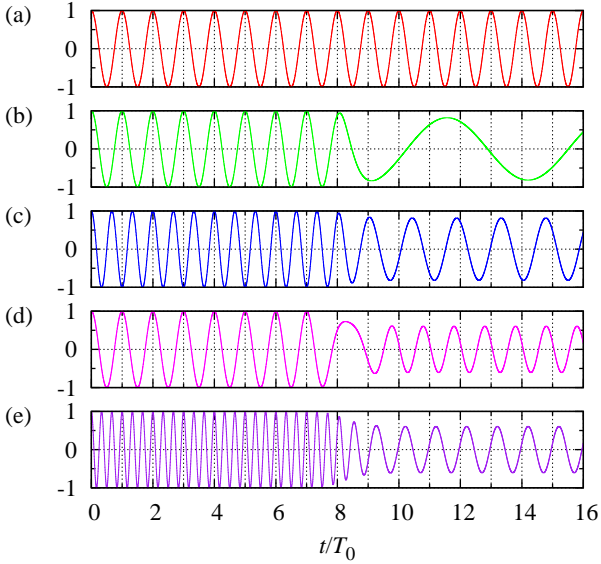}
\caption{(color online). $\text{\rm Re}\,\{a_0(t)\}$ as a function of 
   $t/T_0$ for (a) $\lambda = 0$ and $E_c = 0$, (b) $\lambda = 0.9$ 
   and $E_c = 0$, (c) $\lambda = 0.9$ and $E_c = 0.25$, 
   (d) $\lambda = \sqrt{2}$ and $E_c = 0$, and (e) $\lambda = \sqrt{2}$ 
   and $E_c = 1.0$. Here $N = 4$ and $T_0 = 2\pi\hbar/W_0 = 4\pi$ 
   with $W_0=\hbar\omega/2$. }
\label{Fig1}
\end{figure}

The adiabatic time evolution (\ref{adiabatic-evolution}) is implemented by 
solving the time-dependent Schr\"odinger equation with the forth-order 
Runge-Kutta method on a classical computer. We take $N = 3\sim 6$. 
We assume hat the phase estimation algorithm is implemented very accurately.  
The adiabatic switching-on function $f(t)$ used here is given by 
$f(t) = \frac{1}{2} + \frac{1}{2}\tanh(20\,{t}/{T_R} -10)$.
One may expect it would take a long time for a quantum system to evolve
adiabatically. However, in the case considered here, it takes 
the running time $T_R = 15\ T_0$ to obtain the ground state energy 
with accuracy $\Delta E_0 = |E_0^{\rm exact} - E_0^{\rm num}| \le 10^{-6}$,
where $T_0 \equiv 2\pi/\omega$ is the period of the ground state of 
$H_0$, $E_0^{\rm exact} = \frac{1}{2}\hbar\omega - \frac{\lambda^2}{2m\omega^2}$, 
and $E_0^{\rm num}$ is the numerical value.

\begin{figure}[htbp]
\includegraphics[scale=1.0]{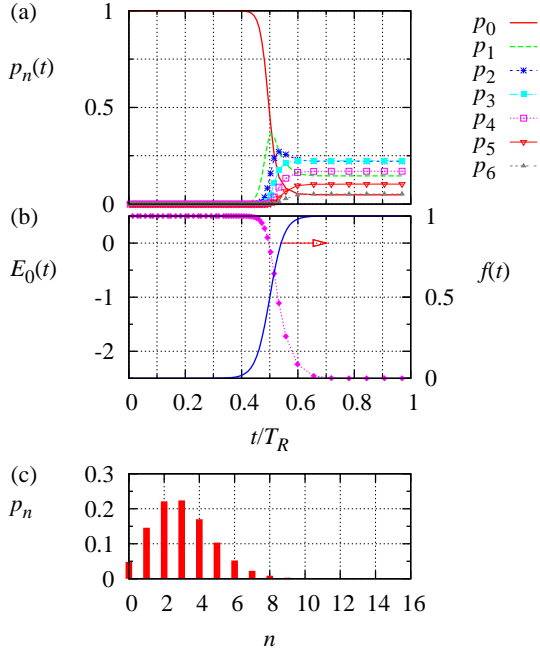}
\caption{(color online). (a) Probability $p_n(t)= |a_n(t)|^2$ of qubits 
   in the computational basis $\ket{n}$, and (b) the instantaneous 
   ground-state energy $E_0(t)$ in the unit of $\hbar\omega$ as a function of 
   the dimensionless time $t/T_R$ for $\lambda = \sqrt{6}$. 
   In (b), $f(t)$ is an adiabatically switching-on function. 
   (c) At $t=T_R$, $p_n$ obeys the Poisson distribution of a coherent state.}
\label{Fig2}
\end{figure}

Fig.~\ref{Fig1} shows how the dynamical phase of the system changes as the
interaction is slowly turned on. In Fig.~\ref{Fig1} (a), $\lambda =0$, and the
oscillation period is $T_0$. Figs.~\ref{Fig1} (b) and (c) show how the constant
energy $E_c$ is used to change the frequency corresponding to the ground state 
energy of an interacting Hamiltonian. In Fig.~\ref{Fig1} (b), $\lambda = 0.9$
and $E_c=0$. So the frequency $E_0/\hbar$ become very low. However, in
Fig.~\ref{Fig1} (c), the constant energy $E_c =1/4$ shifts the frequency so it
can be easily measured. In Figs.~\ref{Fig1} (d) and (e), we take 
$\lambda = \sqrt{2}$ so the exact energy is $E_0= -{1}/{2}$. 
Since the phase estimation algorithm produces 
only the absolute value of energy, $|E_0|$, the constant energy $E_c$ is added 
in (\ref{time_dep_Hamil}) to decide its sign. In Fig.~\ref{Fig1} (d), 
$E_c=0$. At the end of running, the estimated energy is $E_0 = 1/2$. 
So the phase estimation algorithm fails to calculate the exact ground 
state energy $E_0 = -1/2$. However, in Fig.~\ref{Fig1} (e), $E_c = 1$. 
the phase estimation algorithm gives us the energy $1/2$. 
So we know that the exact energy is $E_0 = 1/2 -1 = -1/2$.

\begin{figure}[htbp]
\includegraphics[scale=1.0]{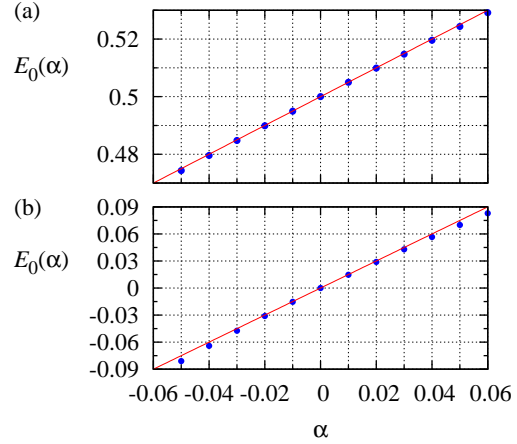}
\caption{(color online). Ground state energy $E_0(\alpha)$ in the unit of $\hbar\omega$ 
   as a function of $\alpha$ for (a) $\lambda = 0$ and (b) $\lambda = 1$. 
   The points are numerical results. The red line in (a) is the plot of 
   $f(\alpha) = \frac{1}{2}(\alpha + 1)$. In (b), the red one is the plot of 
   $g(\alpha) = \frac{3}{2}\alpha $.} 
\label{Fig:HF}
\end{figure}

For any $\lambda$, the ground state of (\ref{DHO}) is a coherent state.
As shown in Fig.~\ref{Fig2}, the probability that qubits are in the number state 
$\ket{n}$  follows a Poisson distribution. So the ground state obtained 
by the quantum simulation  might be called a {\it pseudo-coherent state} 
because it is defined on the truncated 
Hilbert space. It is a collective state of $N$ qubits.

The coherent state is also characterized by the minimum uncertainties 
in $x$ and $p$. Its mean square deviation of $x$, 
$\Delta x^2 = \langle{x^2}\rangle - \langle{x}\rangle^2$ is $1/2$ for 
any $\lambda$. The ground state of~(\ref{DHO}) is displaced from the origin 
to $x=-\lambda$. So $\langle x\rangle = -\lambda$. With the help of 
the Hellmann-Feynman theorem, we calculate $\langle x^2 \rangle$ for 
$\lambda = 0$ and $\lambda = 1$. To this end, the final Hamiltonian is 
modified as $H(t) = H_0 + f(t)(\lambda x + \alpha x^2) + E_c$. 
Fig.~\ref{Fig:HF} shows the ground state energy $E_0(\alpha)$ as a function of 
$\alpha$. The  derivative of $E_0(\alpha)$ at $\alpha = 0$ gives 
us the expectation value of $x$,
$\langle x^2\rangle = \frac{dE_0(\alpha)}{d\alpha}\Bigr|_{\alpha=0}\,.$
As illustrated in Fig.~\ref{Fig:HF},
we have $ \langle{x^2}\rangle = 0.02/0.04 = 1/2$ for $\lambda = 0$. 
Thus $\Delta x^2 = 1/2$.  
For $\lambda = 1$, $\langle x^2 \rangle = 0.03/0.02 = 3/2$. Again we have 
$\Delta x^2 = {3}/{2} - 1 = 1/2$.   

\subsection{Quartic anharmonic oscillator} 

Let us consider an anharmonic oscillator, whose Hamiltonian is given by
\begin{align}
H_0 = \frac{p^2}{2m} + \frac{m\omega^2 x^2}{2}\,,
\quad H_1 = \lambda x^4 \,,
\label{AHO}
\end{align}
where $\lambda >0$ is the coupling constant.
In their seminal paper~\cite{Bender69}, Bender and Wu showed that 
the Rayleigh-Schr\"odinger perturbation theory for (\ref{AHO}) becomes 
divergent for any $\lambda$. Various non-perturbative methods have been 
applied to this simple model.

One can write $H_1 = \frac{\lambda}{4}(a^\dag + a)^4 
= \frac{3\lambda}{4} + \frac{\lambda}{4} V\,$, where 
\begin{align}
V_{mn} = 
  & \sqrt{(n\pm 1)(n \pm 2)(n\pm 3)(n+2 \pm 2)}\, \delta_{m,n\pm 4}\nonumber\\
+ & 2(2n+ 1 \pm 2) \sqrt{(n\pm1)(n+1\pm 1)}\, \delta_{m,n\pm 2}    \nonumber\\
+ & 6n(n+1)\,\delta_{m,n} \,.
\label{AHO_H1}
\end{align}
The matrix of (\ref{AHO_H1}) is more denser than (\ref{DHO_H1}).
So more qubits are used in (\ref{AHO_H1}) in order to get the accurate energy.

\begin{figure}[htbp]
\includegraphics[scale=1.0]{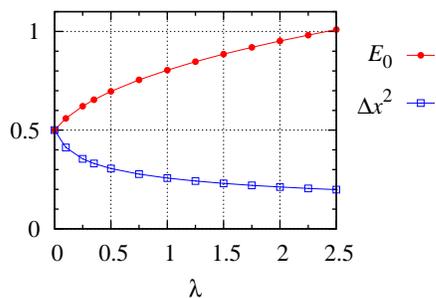}
\caption{(color online). For a quartic anharmonic oscillator, 
   (a) $E_0(\lambda)$ in the unit of $\hbar\omega$ and (b) $\Delta x^2$ 
   as a function of $\lambda$. Here $N=6$ and $T_R=15\ T_0$.}
\label{Fig4}
\end{figure}

Fig.~\ref{Fig4} shows the ground state energy $E_0(\lambda)$ and $\Delta
x^2(\lambda)$ as a function of $\lambda$. For $\lambda = 2.0$ and 
time step $\Delta t= 5.0\times 10^{-5}$, we obtain $E_0 = 0.951\, 568\, 472\,
125$, which is comparable to the best known results 
$E_0 = 0.951\,  568\, 472\, 722 $~\cite{Janke95}. For the calculation of $\Delta
x^2$, we obtain  $E_0(\lambda,\alpha)$ of $H_0 + \lambda H_1 + \alpha x^2$ 
for $\alpha =\pm 0.001$. Thus we have $\langle x^2\rangle \approx  
\left[\, E_0(\lambda,\alpha) -E_0(\lambda,-\alpha)\,\right]/2\alpha$. 

\subsection{Potential scattering model} 

Finally, we consider spinless electrons with a contact potential 
with Hamiltonian 
\begin{align}
H_0 = \sum_{n=1}^{2^N} \epsilon_{n}\, c_n^{\dag} c_n\,,\quad 
H_1 = \frac{g}{2^N}\sum_{n,m} c_n^{\dag} c_{m}\,,
\label{Hamil_PSM}
\end{align}
where $\epsilon_n = (n-1)\Delta$ with level spacing $\Delta$, $c_n^{\dag}$ 
is a creation operator, and $g$ the coupling constant. Although this model 
is very simple and exactly solvable, it contains rich physics~\cite{Kehrein}. 
The naive perturbation theory breaks down no matter small $g$ is. 
For an attractive potential, i.e., $g<0$, the lowest eigenstate of (\ref{Hamil_PSM}) 
becomes a bound state. Also it exhibits the Anderson orthogonality 
catastrophe~\cite{Anderson67} which states that the ground state of $H_0 + H_1$ 
becomes orthogonal to the ground state of $H_0$ in the thermodynamic limit. 

We map the single-particle energy level of $H_0$ to a computational basis,
$\ket{n} = c^\dag_{n}c_n\ket{\rm vac}$, where $\ket{\rm vac}$ is a vacuum state. 
In (\ref{Hamil_PSM}), $H_0$ can be written as a 
diagonal matrix, $(H_0)_{mn} = \epsilon_n\, \delta_{mn}$.
Whereas $H_1$ are given by $(H_1)_{mn} = g/2^{N}$, which is more dense than
(\ref{DHO_H1}) and (\ref{AHO_H1}).
\begin{figure}[htbp]
\includegraphics[scale=1.0]{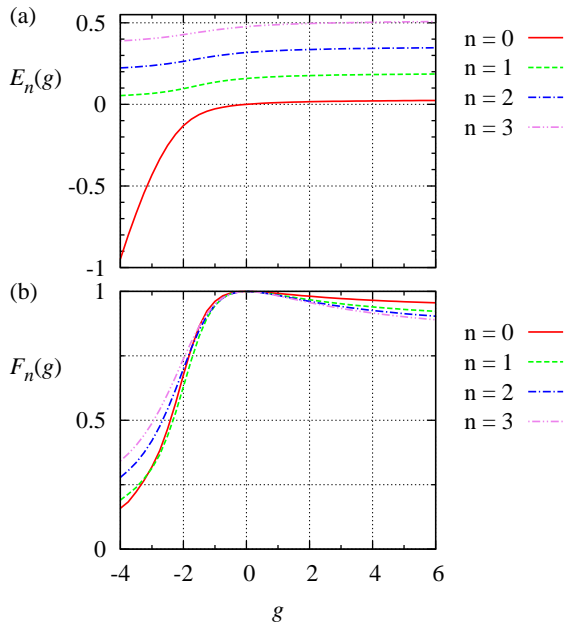}
\caption{(color online). (a) Energy levels $E_n$ (in arbitrary unit) and fidelity $F_n$
  as a function of $g$. Here $N = 6$ and $\Delta=10/64.$ }
\label{Fig:PSM}
\end{figure}

As $g$ is turned on adiabatically, the initial state $\ket{n}$ evolves to 
the final state $\ket{E_n(g)}$. We use the notation $\ket{E_n(0)}= \ket{\epsilon_n} 
= \ket{n}$. Fig.~\ref{Fig:PSM} (a) illustrates the single-particle levels $E_n(g)$.
One see that there is one bound state with negative energy $E_0(g)<0$ for $g<0$, 
but otherwise it is positive. Fig.~\ref{Fig:PSM} (b) shows the fidelity 
$F_n(g) = |\langle{E_n}(g)|{n}\rangle|^2$ as a function of $g$. Surprisingly, 
it is also calculated with the help of the Hellmann-Feynman theorem. One can
rewrite $F_n(g) = \bra{E_n(g)}{\cal O}\ket{E_n(g)}$ with ${\cal O} = \ket{n}\bra{n}$.
As shown in Fig.~\ref{Fig:PSM} (b), the fidelity decrease more rapidly for $g<0$
than for $g>0$. One can see that even single-particle levels 
for $g=0$ and $g<0$ become orthogonal. It is interesting that the fidelity 
between the interacting and non-interacting many-body ground states 
can be obtained from all the information of single-particle 
levels~\cite{Ohtaka90}.

\section{conclusions}

In conclusion, we have proposed a new method to find the ground state 
energy by adiabatically turning on an interaction. The expectation 
values of an observable has been obtained by switching on a modified 
interaction which contains an observable and by applying 
the Hellmann-Feynman theorem. Our method has been successfully 
tested by solving three quantum systems. We expect that our method could be
applied to the simulation of more interesting quantum systems.

Finally, let us discuss the limits of our method. 
Our method is based on the combination of adiabatic quantum 
computation and the phase estimation algorithm. So, the computational resources needed to 
implement our method is approximately equal to the sum of those involved in adiabatic 
quantum computation and the phase estimation algorithm.
The running time of the adiabatic evolution increases if the gap between the energy levels 
decreases. However, it is expected that the quantum Zeno effect~\cite{Misra77} 
might release this limitation.
A quantum state after applying a quantum phase estimation algorithm 
is approximately given by $\ket{\Psi(t)} \approx a_0\ket{E_0}_S\,\ket{\omega_0}_I  
+ a_1\ket{E_1}_S\,\ket{\omega_1}_I$ where $|a_1|^2 = 1 -\epsilon$ and $|a_1|^2 = \epsilon$ for small 
$\epsilon$ and subscripts ``S" and ``I" refer to the system and 
the index qubits, respectively. The measurement on the index qubits gives us
$\ket{\Psi(t)} = \ket{E_0}_S\,\ket{\omega_0}_I$ with high probability.
The frequent applications of a quantum phase estimation algorithm 
and measurement on the index qubits could accelerate an adiabatic evolution.
This will be investigated in a future study.

\acknowledgments
S.O. thanks A. Buchleitner for helpful comments.

\end{document}